\documentclass[twoside,11pt]{article}

\usepackage{jmlr2e}
\usepackage{caption}
\usepackage{multicol}
\usepackage{amsmath,amssymb,amsfonts}
\usepackage{graphicx}
\usepackage{textcomp}
\usepackage{xcolor}
\usepackage{tabularx}
\ShortHeadings{Spectral Enhancement of Lossy Audio Sequences}{Deshpande and Abichandani}
\firstpageno{1}
\begin{document}
% \title{Audio Spectral Enhancement: A Low Latency Reconstruction Approach for Long Sequences of Lossy Audio}
\title{Audio Spectral Enhancement: Leveraging Autoencoders for Low Latency Reconstruction of Long, Lossy Audio Sequences}

\author{\name Darshan Deshpande \email darshan.deshpande\_student@thadomal.org \\
       \addr Department of Computer Engineering\\
       Thadomal Shahani Engineering College \\
       Mumbai University \\
       Mumbai, India.
    %   Seattle, WA 98195-4322, USA
       \AND
       \name Harshavardhan Abichandani \email \text{\footnotesize harshavardhan.abichandani\_student@thadomal.org} \\
       \addr Department of Computer Engineering\\
       Thadomal Shahani Engineering College \\
       Mumbai University \\
       Mumbai, India.}
    %   Berkeley, CA 94720-1776, USA}

\editor{}

\maketitle

\begin{abstract}%   <- trailing '%' for backward compatibility of .sty file
\noindent With active research in audio compression techniques yielding substantial breakthroughs, spectral reconstruction of low-quality audio waves remains a less indulged topic. In this paper, we propose a novel approach for reconstructing higher frequencies from considerably longer sequences of low-quality MP3 audio waves. Our technique involves inpainting audio spectrograms with residually stacked autoencoder blocks by manipulating individual amplitude and phase values in relation to perceptual differences. Our architecture presents several bottlenecks while preserving the spectral structure of the audio wave via skip-connections. We also compare several task metrics and demonstrate our visual guide to loss selection. Moreover, we show how to leverage differential quantization techniques to reduce the initial model size by more than half while simultaneously reducing inference time, which is crucial in real-world applications.

\end{abstract}


\section{Introduction}
\noindent Data compression is utilised in a variety of ways, from streaming music to transferring text files, but it has had the greatest impact on the audio and image domains. Among all audio compression algorithms, MP3 is a well-known codec, primarily because it drastically decreases the file size while having little effect on audio quality. It uses a perceptual model which removes frequencies that are imperceptible to humans, usually involving the high frequencies. The way it does this is by changing the input from the time domain to the frequency domain by using FFT and subsequently determining the threshold values of these frequencies. After this, Huffman encoding is applied to further compress the file by assigning lower number of bits to the frequently occurring frequencies. MP3 is a lossy compression format since it removes information from the audio that is difficult to recover. The loss of information is not perceptible by humans at high enough bitrates, but at low bitrates artifacts like pre-echo are noticeable.

Some techniques have been established previously, in which the algorithm would iteratively estimate the phase source and reconstruct the audio \citep{algo1:10, algo2:12, algo3:14}. These techniques of estimating the phase show considerable improvement in the SSNR and PESQ score, but all of them ignore other characteristics of the audio like the magnitude and power spectrum \citep{Wang:19}. In recent times machine learning methods have been used to outperform these algorithmic techniques. CNN models were utilised, with the real and imaginary spectrograms of the audio serving as separate input channels for the model \citep{Szu:17}. The resultant spectrograms were then used for the reconstruction of the signal, this type of reconstruction outperformed the algorithmic approaches.

Convolutional Recurrent Neural Network (CRN) model proposed by \citep{Wang:19} had a substantially higher STOI and PESQ value compared to the CNN based approach mentioned above \citep{Szu:17}. Later, \citep{Kuleshov:17} proposed a technique in which time-series data is fed into an autoencoder with residual connections, which reconstructs the high resolution audio with a sampling rate higher than the low resolution audio. This model was inspired by the image super resolution paper \citep{Dong:16} and demonstrates an improvement of about 5dB in some circumstances \citep{Kuleshov:17}. Resnet and Unets \citep{Sulun:21} have also been tested for the task of bandwidth extension. Multiple low pass filters were utilised to enhance the data, allowing these models to generalize effectively on new unseen filters \citep{Sulun:21}. The code for the sections that follow can be found at: \url{https://github.com/DarshanDeshpande/audio-spectral-enhancement}.

\section{Processing the Data}
\noindent The usage of extended temporal sequences is the fundamental focus of our research (above 100,000 data points per training sample). Because of the limited input size and repeated sweeps of segments through the decoder network that are necessary to rebuild lengthier audio sequences, upscaling with some previous reconstruction algorithms like \citep{Segan:17} is significantly slower. Although parallelism helps speed up the process, larger audio files still take a long time to process. To help mitigate this major constraint, one of the most practical solutions is to analyse and reconstruct larger audio sequences in a single inference which is what we use for our proposed model. This section provides details about the creation of training data and the structuring of the pipeline.

\subsection{Dataset}
\noindent We use the GTZAN \citep{GTZAN:02} dataset as a base for our experiments. It includes 1,000 WAV tracks, each of which lasts 30 seconds and are sampled at 22050 Hz. Each of these songs was divided into three parts, augmenting the total number of songs in the dataset to 3,000. Using the LAME encoder \citep{lame-encoder} and the FFMPEG library\citep{FFmpeg:06}, all 3,000 mono 16bit WAV files were first converted to MP3 files at a constant bitrate of 128kbps. We then downscaled the 128kbps audio files to 32kbps audio files using the same LAME encoder and the FFMPEG library\citep{FFmpeg:06}. We have used $85\%$ of the data for training, $8\%$ data for validation and $7\%$ data for testing. 

\subsection{Loading the Data}
\noindent To accurately capture the differences between high and low bitrate audio, the data was preprocessed using the Short Time Fourier Transform (STFT). The STFT function returns a tensor of complex STFT values containing information about the signal's phase and amplitude, which was then separated into real and imaginary parts and stacked on top of each other. Stacking was done because the real and imaginary values are directly mappable to the new real and imaginary values, respectively.

We experimented with various frame length and hop size configurations to retrieve the most data while minimizing processing time. We found that a Hanning window length of 1023, hop size of 248, a frame length of 1024 without padding or centering provided the best results. This configuration gives us 400 frames and 512 frequency bins for every sample of 100,000 data points. The hop size should be small and the frame length should be large to achieve accurate inversion. Lowering the hop size to a very small value increases the time resolution but it is very computationally heavy. This leads to increased inference time and as a result, our setup finds a good compromise between the two.
% The data was scaled by the factor of $2/512$ where the denominator is half of the frame length.

\section{Model Creation}
\noindent Popular purely convolutional networks such as ResNets \citep{He:16} and UNets \citep{Unet:15}, rely heavily on residual connections and lower dimensional projections. When training data is scarce and fast inference is of high priority, our experiments revealed that these large models are significantly less efficient. In our study, we found that 2-D Convolutional UNets \citep{Unet:15} consistently produced poor latent representations, drawing more attention to the background noise instead of the missing frequencies. Furthermore, 1-D UNet based models like Wave-U-Net \citep{WaveUnet:18}, Temporal FiLM \citep{Tfilm:19} and Audio-SR-Net \citep{Sulun:21} have only been tested for sequences with 10,000 or fewer data points per sample. We discovered that 1-D convolutions applied to training samples temporally ten times that size are extremely inefficient and computationally costly to train. The subsections that follow detail the experiments and model designs that were employed.

\subsection{Model Architecture} \label{model-architecture}
\noindent Since one-dimensional models are unusable as discussed above, it is more cost and time effective to directly convolve on the stacked STFT output. Previously mentioned 2-D UNet architecture \citep{Unet:15} was designed with a fixed positive range of 0 to 1 for images in mind. Since we had to scale our data in order to feed it to the 2-D UNet model \citep{Unet:15}, we had two options for dealing with the issue of negative values in the scaled data: the first was to change the range of our data to 0 to 1, and the second was to use activation functions like \textit{tanh}, which support negative values. In comparison to the first method, which resulted in insufficient weighting of negative values, our comparative trials demonstrated that the second strategy resulted in significantly improved spectral structure retention. Both the PReLU \citep{He:15} and \textit{tanh} functions work well with the 2-D U-Net \citep{Unet:15}, but they quickly overfit due to the large number of trainable parameters and small dataset size, resulting in an extremely saturated spectral plot and poor audio output.

\begin{figure*}
% \begin{multicols}{1}
    \centering
    \vspace{.5mm}
    \includegraphics[width=100mm]{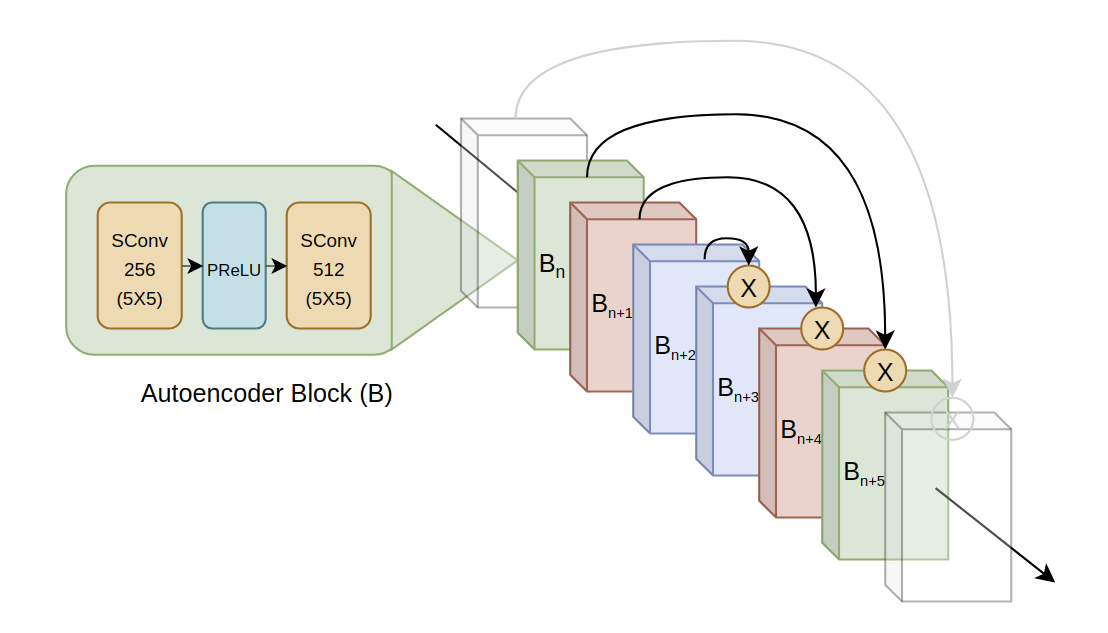}\vspace{.5mm}
    % \includegraphics[scale=0.5]{assets/Architecture3D.png}\par
% \end{multicols}
\caption{A single Depthwise Separable Convolution Block(B) and it's arrangement as a residually connected autoencoder network}
\label{architecture}
\end{figure*}

Our proposed model consists of a two-layer 2-D depthwise separable convolutional \citep{Laurent:14} block, as seen in figure  \ref{architecture}. This architecture functions as a tiny autoencoder unit with a 256-dimensional latent space. Each block uses a PReLU \citep{He:15} activation which prevents the gradients from exploding while performing significantly better than \textit{tanh} function in our experiments. With PReLU \citep{He:15}
and $\alpha$ parameter initialization of zero, the model attains more freedom to explore negative values which is not possible with the standard ReLU or sigmoid activation functions. This activation function is sandwiched between separable convolutional layers \citep{Laurent:14}, each with a filter size of $5 \times 5$ and step size of 1. Our model uses constant padding which persists the input rows in the output, along with a randomly uniform initialization in accordance with the initialization proposed by \citep{He:15}. The use of separable convolutions \citep{Laurent:14} rather than general spatial convolutions stems from their low likelihood of overfitting \citep{Laurent:14}, as demonstrated by our experiments with considerably higher trainable parameters and various custom experiments that are shown in table \ref{metrictable}. Each block can be stacked \textit{N} times with skip connections as shown in figure  \ref{architecture} to provide effective protection against the vanishing gradient problem \citep{He:16}. A significant observation is that with the increase of \textit{N}, the coherence of the output audio decreases despite the phase and power spectra remaining consistent. Through our experiments, we could also conclude that a model consisting upto seven autoencoder blocks does not lead to overfitting on our experiments with our augmented GTZAN dataset \citep{GTZAN:02}. This explains the robust architectural design of the configuration primarily due to the use of separable convolutions \citep{Laurent:14}. 

% \begin{figure*}
% \begin{center}
% \hspace*{-1.5cm}
% \begin{tabular}{c c c c}
%     \includegraphics[width=40mm]{assets/Low.png}&
%     \includegraphics[width=40mm]{assets/High.png} &
%     \includegraphics[width=40mm]{assets/R2only.png}&
%     \includegraphics[width=40mm]{assets/SSIMonly.png}& 
%     \includegraphics[width=40mm]{assets/MSEonly.png} &
%     \includegraphics[width=40mm]{assets/MAEonly.png} &
%     \includegraphics[width=40mm]{assets/SPECCONVonly.png} &
%     \includegraphics[width=40mm]{assets/BCE.png}
% \end{tabular}
% \end{center}
% \label{loss-comparison}
% \end{figure*}

\begin{figure*}
    \centering
    \includegraphics[width=\textwidth]{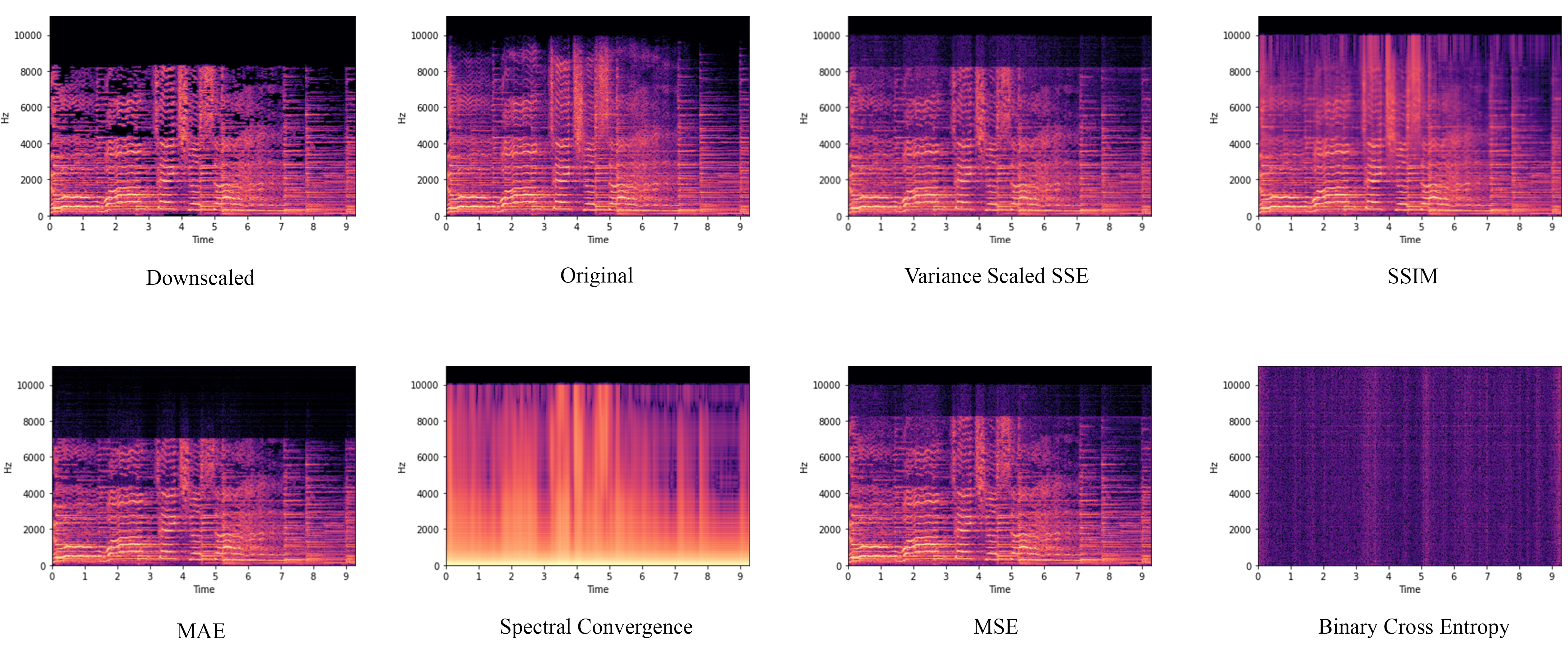}
    \captionof{figure}{Outputs of different loss functions}
    \label{loss-comparison}
\end{figure*}

\subsection{Training Losses}\label{training-losses}
\noindent Our experimentation with loss functions involved permuting between six losses and testing their combinations. The figure \ref{loss-comparison} provides a summary of the individual metric tests. Considering these effects, we propose a weighted combination of two losses for our training process. First is a modified version of the sum of squared residuals (SSE) which is scaled by the variance of the ground truth label.

\begin{equation}
\text{Pixel Loss}(y_i, \hat{y_i}) = \displaystyle \frac{\sum_{i=1}^n{(y_i - \hat{y_i}})^2}{\sum_{i=1}^n{(y_i - \bar{y_i}})^2}
\end{equation}

This loss function provides a much better spectral structure as compared to the standard Mean Squared Error (MSE) for unscaled data or Kullback-Leibler Divergence (KDL) for scaled data which produced faded textures. Similarly, Mean Absolute Error (MAE) produced results that neglected the higher frequencies altogether because of it's inherent nature of ignoring outliers. Some commonly employed spectral reconstruction losses, such as Log  Spectral Loss \citep{Hu:08}, resulted in oversaturated low-frequency regions that were not affected by compression, to begin with, as seen in figure  \ref{loss-comparison}.

The second loss that we use is the Structural Similarity Index (SSIM) \citep{Wang:04}. The idea for this function is derived from its large applicability in domains such as image reconstruction and image super-resolution. SSIM highly influences the spectral luminosity and contrast of the output spectrogram and ensures that the higher frequency intensity ranges are equally stressed on unlike the Log Spectral Loss \citep{Hu:08}

\begin{equation}
    \text{SSIM}(y_i,\hat{y_i}) = \displaystyle \frac{(2\mu_{y_i}\mu_{\hat{y_i}} + C_1)(2\sigma_{y_i \hat{y_i}} + C_2)}{(\mu_{y_i}^2 + \mu_{\hat{y_i}}^2 + C_1)(\sigma_{y_i}^2 + \sigma_{\hat{y_i}}^2 + C_2)}
\end{equation}

Here, the SSIM loss is calculated on the magnitude of the output. $C_1=(k_1L)^2$ and $C_2=(k_2L)^2$ are constants for stabilization determined by the dynamic pixel value ranges, \textit{L}, and pre-determined constants $k_1$, $k_2$. Changing the value of $k_1$ even slightly, changes the luminosity by a large factor and hence this value is most optimal if set between 0.01 and 0.005. Additionally, altering the value of $k_2$ gives a better definition to the output spectral plot at the cost of low intensity of higher frequencies. Thus, the appropriate value of $k_2$ can range from 0.01 to 0.03. The value of \textit{L} should be set to less than $1/2$ of the actual pixel range to allow for better luminosity. \citep{Wang:04} recommend that the technique should be applied locally for better results. Our exhaustive experiments conclude that applying the loss using a smaller convolutional window size of $3 \times 3$ improves the structural detail of the output and prevents checkerboard patterns.

% \begin{figure*}
%     \centering
%     \includegraphics[width=\textwidth]{assets/alllosses.png}
%     \caption{Caption}
%     \label{fig:my_label}
% \end{figure*}

\subsection{Training}
\noindent The training process involves using the Adam \citep{Kingma:14} optimizer and takes approximately 200 epochs to converge with a learning rate of $ 1 \times 10^3$ and a mini batch size of 32. Other popular optimizers, such as mini-batch SGD and RMSProp, are prone to get trapped in local minima, resulting in insufficient loss landscape mapping. The training time heavily depends on the model configuration, number of parameters and the value of \textit{N} mentioned in subsection \ref{model-architecture}. Additionally, a cyclical learning rate with a decay factor of $5 \times 10^{-4}$ can assist in providing a smoother descent. 

At every step, the losses are weighted as follows:

\begin{equation}
\text{Total Loss} = \displaystyle \text{Pixel Loss} + 0.5 \times \text{SSIM}
\end{equation}

This weighting is used to counteract SSIM's tendency to dominate the spectrogram by filling it with unnecessary high-intensity pixels. This prevents the spectrogram from being overly saturated and producing a noisy audio output.

\subsection{Model Quantization}
\noindent Since the primary goal of this research is to provide a low latency solution for spectral mapping, we adopt quantization techniques \citep{Jacob:18} to shrink the model size as well as decrease inference times. We used a dynamic 8-bit precision quantization to achieve a compression of up to 52.15\%, resulting in a compressed size of 956KB per block. In this quantization technique, the activations are dynamically quantized and un-quantized between 8-bit and float precisions during and after training respectively, while the weights are stored with 8-bit precision. The differences in spectrograms after compression is minuscule, with 16-bit float quantization techniques retaining the complete spectral structure and the 8-bit integer quantization techniques altering the intensity of the higher frequencies just slightly, while ensuring the persistence of the quality of the audio output.

% \begin{center}
% \begin{tabular}{c c c}
%     \includegraphics[width=50mm]{assets/MODELOGQUANT.png} &  \includegraphics[width=50mm]{assets/FLOAT16QUANT.png} &
%     \includegraphics[width=50mm]{assets/INT32QUANT.png}\\
% \end{tabular}
% \captionof{figure}{Left to right: }
% \label{}
% \end{center}

\section{Evaluation}

\noindent The evaluation section is divided into two parts. First, the spectrograms of high, low and approximated audio signals are compared and finally, some of the most common objective and perceptual based metrics are used to evaluate the audio signals.

\subsection{Spectrograms}
\begin{figure}
\setlength\fboxsep{0pt}
\setlength\fboxrule{0.25pt}
\includegraphics[width=\textwidth]{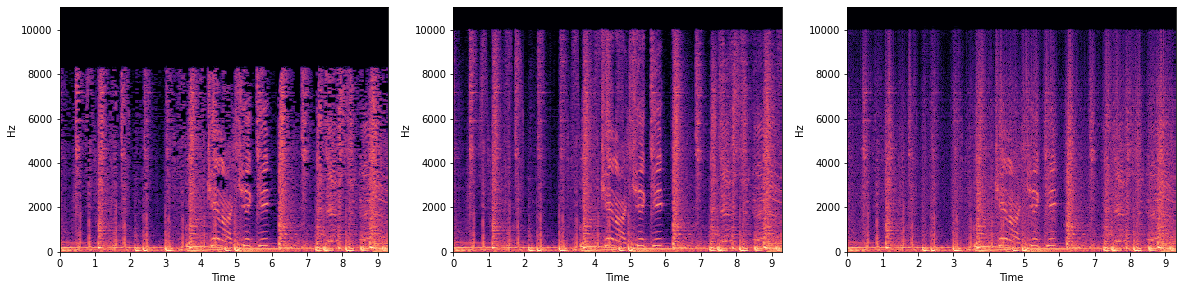} 
\caption{From left to right: low bitrate audio spectrogram, high bitrate audio spectrogram and reconstructed audio spectrogram}
\label{spectrograms}
\end{figure}

\begin{figure}
\centering
\begin{multicols}{3}
    \setlength\fboxsep{0pt}
    \setlength\fboxrule{0pt}
    \includegraphics[width=45mm]{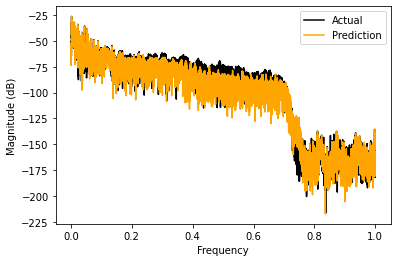}\par
    \includegraphics[width=45mm]{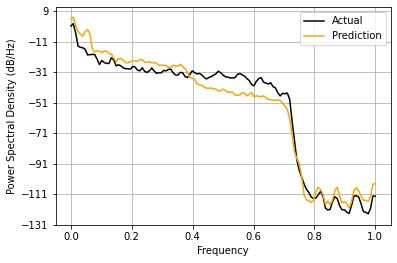}\par 
    \includegraphics[width=45mm]{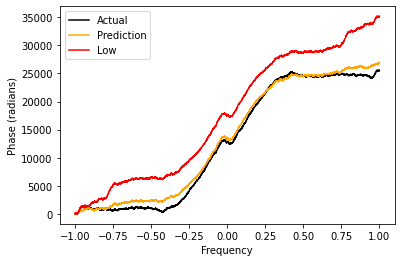}\par 
\end{multicols}
\caption{From left to right: magnitude spectrogram, power spectrogram and two-sided phase spectrogram}
\label{phasemagpow}
\end{figure}

\noindent Inverse STFT of the approximated signal was taken and then its frequency, phase, magnitude and power spectrograms were plotted against the high bitrate audio signal as shown in figure  \ref{phasemagpow}. Figure \ref{spectrograms} shows that the reconstructed spectrogram has kept the structure of higher frequencies. Figure \ref{phasemagpow} shows a two-sided phase spectrogram that displays how the reconstructed phase differs just slightly from the high bitrate audio.

\subsection{Metrics}
\noindent The performance of the reconstructed audio signal $\hat{y}$ in comparison to the high bitrate audio signal $y$ was evaluated using the following metrics. 
\begin{enumerate}
    \item Signal to Noise Ratio (SNR) \\
    \begin{equation}
        \text{SNR}(y, \hat{y}) = \displaystyle 10\cdot\log_{10}\frac{||y||_{2}^{2}}{||y - \hat{y}||_{2}^{2}}
    \end{equation}
    Signal to Noise ratio is the most widely used metric for analyzing the clarity of the audio signals. The higher the value of the ratio, the better is the audio quality.
    \item Perceptual Evaluation of Speech Quality (PESQ) \\
    It is calculated by taking the linear combination of the average disturbance values $D_{ind}$ and the average asymmetrical disturbance values $A_{ind}$ \citep{Hu:08, Rix:01}.
    \begin{equation}
        \text{PESQ}(y, \hat{y}) = a_{0} + a_{1}D_{ind} + a_{2}A{ind} 
    \end{equation}
    We have taken all the default values of $a_{0}$, $a_{1}$, and $a_{2}$ and used a wideband configuration with the sampling frequency of 16000Hz as mentioned in \citep{pesq-github:21}.
    \item Log-Spectral Distance (LSD) \\
    \begin{equation}
        \text{LSD}(y, \hat{y}) = \frac{1}{L}\sum\limits_{l=1}^{L}\sqrt{\frac{1}{K}\sum\limits_{k=1}^{K}(S(k) - \hat{S}(k))^2}
    \end{equation}
    here $S$ and $\hat{S}$ are calculated by first taking the STFT of $y$ and $\hat{y}$ respectively \citep{distance-measures-Gray:76}. For the STFT, a window length of 2048 was chosen, followed by the log of the square of absolute values of the STFT. Here $K$ and $L$ represents the frequencies and the total number of frames respectively.
    
    \item Short-Time Objective Intelligibility (STOI) \\
    STOI was calculated based on the parameters mentioned in \citep{Taal:10, Taal:11} with the high bitrate audio, the reconstructed audio and the sampling rate equal to $44100Hz$ as the inputs to the function. 
\end{enumerate}

\begin{table}[ht]
\captionof{table}{Inference time in milliseconds and metric comparison of different models.}
\begin{center}
    \begin{tabular}{|c|c|c|c|c|c|c|}
    \hline
    & \textbf{Audio-SR-Net} & \textbf{FFmpeg} & \textbf{TFiLM} & \textbf{WaveNet} & \textbf{N = 1} & \textbf{N = 5} \\
    & (27.7ms) & & (75.9ms) & (86.9ms) & (7.74ms) &(20.8ms) \\
    \hline
    \textbf{SNR} & 16.7353 & 18.0077 & 9.9723 & 13.6910 & \textbf{21.2532} & 20.8868 \\
    \textbf{PESQ} & 2.6296 & 4.3039 & 3.6352 & 4.1743 & \textbf{4.3599} & 4.1197 \\
    \textbf{LSD} & 2.3340 & 2.6845 & 2.6497 & 3.1490 & \textbf{1.5909} & 1.8152  \\
    \textbf{STOI} & \textbf{0.9996} & 0.9913 & 0.9635 & 0.9881 & 0.9910 & 0.9894 \\
    \hline
\end{tabular}
\end{center}
\label{metrictable}
\end{table}

In table \ref{metrictable}, a single autoencoder block (N = 1) outperforms the majority of the models compared. Even though it contains six times the number of trainable parameters as a single block, the five block autoencoder (N = 5) shows a slight variation in metric values and does not overfit. This demonstrates that a small sized model is adequate to generalize and is sufficiently robust that, when the size is increased, it shows no major evidence of overfitting when compared to the other techniques.

\section{Conclusion and Improvements}
\noindent In this paper, we propose a novel architecture involving nested, residually connected autoencoder blocks for the spectral reconstruction of monaural audio samples. The preprocessing pipeline involves the conversion of a higher, 128Kbps audio wave to a lower, 32Kbps audio using FFMPEG \citep{FFmpeg:06} whose reconstruction was then attempted using the residual network. We compare our method against FFMPEG \citep{FFmpeg:06} based reconstruction and a variety of preexisting and new machine learning methods using popular perceptual metrics such as PESQ \citep{Rix:01} and LSD \citep{distance-measures-Gray:76}. Furthermore, we discuss how the proposed model can be pruned to a minimum size of 956KB (52.15\%) per autoencoder block, resulting in perceptually comparable inferences.

Despite the benefits of the architecture and techniques used, the output wave sometimes tends to be slightly incoherent with the source. Furthermore, due to a lack of open sourced trainable data, the model could only be trained on 3,000 samples, resulting in poor generalisation of the silent regions in the audio spectra. Additionally, Generative Adversarial Networks (GANs) \citep{GAN:14} and Audio Transformers \citep{audio-spec-transformer}, and the use of coherence-based comparative metrics are two noteworthy approaches to consider for future trials.

\vskip 0.2in
\bibliography{citations}

\begin{thebibliography}{29}
\providecommand{\natexlab}[1]{#1}
\providecommand{\url}[1]{\texttt{#1}}
\expandafter\ifx\csname urlstyle\endcsname\relax
  \providecommand{\doi}[1]{doi: #1}\else
  \providecommand{\doi}{doi: \begingroup \urlstyle{rm}\Url}\fi

\bibitem[Birnbaum et~al.(2019)Birnbaum, Kuleshov, Enam, Koh, and
  Ermon]{Tfilm:19}
Sawyer Birnbaum, Volodymyr Kuleshov, S.~Enam, Pang~Wei Koh, and S.~Ermon.
\newblock Temporal film: Capturing long-range sequence dependencies with
  feature-wise modulations.
\newblock In \emph{NeurIPS}, 2019.

\bibitem[Cheng(1998)]{lame-encoder}
Mike Cheng.
\newblock {LAME} {MP3} {Encoder}, 1998.
\newblock URL \url{https://lame.sourceforge.io/}.

\bibitem[Dong et~al.(2016)Dong, Loy, He, and Tang]{Dong:16}
Chao Dong, Chen~Change Loy, Kaiming He, and Xiaoou Tang.
\newblock Image super-resolution using deep convolutional networks.
\newblock \emph{IEEE Transactions on Pattern Analysis and Machine
  Intelligence}, 38\penalty0 (2):\penalty0 295--307, 2016.
\newblock \doi{10.1109/TPAMI.2015.2439281}.

\bibitem[Fu et~al.(2017)Fu, Hu, Tsao, and Lu]{Szu:17}
Szu-Wei Fu, Ting-yao Hu, Yu~Tsao, and Xugang Lu.
\newblock Complex spectrogram enhancement by convolutional neural network with
  multi-metrics learning.
\newblock In \emph{2017 IEEE 27th International Workshop on Machine Learning
  for Signal Processing (MLSP)}, pages 1--6, 2017.
\newblock \doi{10.1109/MLSP.2017.8168119}.

\bibitem[Gong et~al.(2021)Gong, Chung, and Glass]{audio-spec-transformer}
Yuan Gong, Yu{-}An Chung, and James~R. Glass.
\newblock {AST:} audio spectrogram transformer.
\newblock \emph{CoRR}, abs/2104.01778, 2021.
\newblock URL \url{https://arxiv.org/abs/2104.01778}.

\bibitem[Goodfellow et~al.(2014)Goodfellow, Pouget-Abadie, Mirza, Xu,
  Warde-Farley, Ozair, Courville, and Bengio]{GAN:14}
Ian Goodfellow, Jean Pouget-Abadie, Mehdi Mirza, Bing Xu, David Warde-Farley,
  Sherjil Ozair, Aaron Courville, and Yoshua Bengio.
\newblock Generative adversarial nets.
\newblock \emph{Advances in neural information processing systems}, 27, 2014.

\bibitem[Gray and Markel(1976)]{distance-measures-Gray:76}
A.~Gray and J.~Markel.
\newblock Distance measures for speech processing.
\newblock \emph{IEEE Transactions on Acoustics, Speech, and Signal Processing},
  24\penalty0 (5):\penalty0 380--391, 1976.
\newblock \doi{10.1109/TASSP.1976.1162849}.

\bibitem[Gunawan and Sen(2010)]{algo1:10}
David Gunawan and D.~Sen.
\newblock Iterative phase estimation for the synthesis of separated sources
  from single-channel mixtures.
\newblock \emph{IEEE Signal Processing Letters}, 17\penalty0 (5):\penalty0
  421--424, 2010.
\newblock \doi{10.1109/LSP.2010.2042530}.

\bibitem[He et~al.(2015)He, Zhang, Ren, and Sun]{He:15}
Kaiming He, Xiangyu Zhang, Shaoqing Ren, and Jian Sun.
\newblock Delving deep into rectifiers: Surpassing human-level performance on
  imagenet classification.
\newblock In \emph{Proceedings of the IEEE international conference on computer
  vision}, pages 1026--1034, 2015.

\bibitem[He et~al.(2016)He, Zhang, Ren, and Sun]{He:16}
Kaiming He, Xiangyu Zhang, Shaoqing Ren, and Jian Sun.
\newblock Deep residual learning for image recognition.
\newblock In \emph{Proceedings of the IEEE conference on computer vision and
  pattern recognition}, pages 770--778, 2016.

\bibitem[Hu and Loizou(2008)]{Hu:08}
Yi~Hu and Philipos~C. Loizou.
\newblock Evaluation of objective quality measures for speech enhancement.
\newblock \emph{IEEE Transactions on Audio, Speech, and Language Processing},
  16\penalty0 (1):\penalty0 229--238, 2008.
\newblock \doi{10.1109/TASL.2007.911054}.

\bibitem[Jacob et~al.(2018)Jacob, Kligys, Chen, Zhu, Tang, Howard, Adam, and
  Kalenichenko]{Jacob:18}
Benoit Jacob, Skirmantas Kligys, Bo~Chen, Menglong Zhu, Matthew Tang, Andrew
  Howard, Hartwig Adam, and Dmitry Kalenichenko.
\newblock Quantization and training of neural networks for efficient
  integer-arithmetic-only inference.
\newblock In \emph{Proceedings of the IEEE conference on computer vision and
  pattern recognition}, pages 2704--2713, 2018.

\bibitem[Kingma and Ba(2014)]{Kingma:14}
Diederik~P Kingma and Jimmy Ba.
\newblock Adam: A method for stochastic optimization.
\newblock \emph{arXiv preprint arXiv:1412.6980}, 2014.

\bibitem[Krawczyk and Gerkmann(2014)]{algo3:14}
Martin Krawczyk and Timo Gerkmann.
\newblock Stft phase reconstruction in voiced speech for an improved
  single-channel speech enhancement.
\newblock \emph{IEEE/ACM Transactions on Audio, Speech, and Language
  Processing}, 22\penalty0 (12):\penalty0 1931--1940, 2014.
\newblock \doi{10.1109/TASLP.2014.2354236}.

\bibitem[Kuleshov et~al.(2017)Kuleshov, Enam, and Ermon]{Kuleshov:17}
Volodymyr Kuleshov, S.~Zayd Enam, and Stefano Ermon.
\newblock Audio super resolution using neural networks, 2017.

\bibitem[Mowlaee et~al.(2012)Mowlaee, Saeidi, and Martin]{algo2:12}
Pejman Mowlaee, Rahim Saeidi, and Rainer Martin.
\newblock Phase estimation for signal reconstruction in single-channel source
  separation.
\newblock In \emph{Thirteenth Annual Conference of the International Speech
  Communication Association}, 2012.

\bibitem[Pascual et~al.(2017)Pascual, Bonafonte, and Serrà]{Segan:17}
Santiago Pascual, Antonio Bonafonte, and Joan Serrà.
\newblock Segan: Speech enhancement generative adversarial network.
\newblock In \emph{Proc. Interspeech 2017}, pages 3642--3646, 2017.
\newblock \doi{10.21437/Interspeech.2017-1428}.
\newblock URL \url{http://dx.doi.org/10.21437/Interspeech.2017-1428}.

\bibitem[Rix et~al.(2001)Rix, Beerends, Hollier, and Hekstra]{Rix:01}
A.W. Rix, J.G. Beerends, M.P. Hollier, and A.P. Hekstra.
\newblock Perceptual evaluation of speech quality (pesq)-a new method for
  speech quality assessment of telephone networks and codecs.
\newblock In \emph{2001 IEEE International Conference on Acoustics, Speech, and
  Signal Processing. Proceedings (Cat. No.01CH37221)}, volume~2, pages 749--752
  vol.2, 2001.
\newblock \doi{10.1109/ICASSP.2001.941023}.

\bibitem[Ronneberger et~al.(2015)Ronneberger, Fischer, and Brox]{Unet:15}
Olaf Ronneberger, Philipp Fischer, and Thomas Brox.
\newblock U-net: Convolutional networks for biomedical image segmentation.
\newblock In Nassir Navab, Joachim Hornegger, William~M. Wells, and
  Alejandro~F. Frangi, editors, \emph{Medical Image Computing and
  Computer-Assisted Intervention -- MICCAI 2015}, pages 234--241, Cham, 2015.
  Springer International Publishing.
\newblock ISBN 978-3-319-24574-4.

\bibitem[SIfre and Mallat(2014)]{Laurent:14}
Laurent SIfre and Stéphane Mallat.
\newblock Rigid-motion scattering for texture classification, 2014.

\bibitem[Stoller et~al.(2018)Stoller, Ewert, and Dixon]{WaveUnet:18}
D.~Stoller, S.~Ewert, and S.~Dixon.
\newblock Wave-u-net: A multi-scale neural network for end-to-end audio source
  separation.
\newblock \emph{ArXiv}, abs/1806.03185, 2018.

\bibitem[Sulun and Davies(2021)]{Sulun:21}
Serkan Sulun and Matthew E.~P. Davies.
\newblock On filter generalization for music bandwidth extension using deep
  neural networks.
\newblock \emph{IEEE Journal of Selected Topics in Signal Processing},
  15\penalty0 (1):\penalty0 132--142, 2021.
\newblock \doi{10.1109/JSTSP.2020.3037485}.

\bibitem[Taal et~al.(2010)Taal, Hendriks, Heusdens, and Jensen]{Taal:10}
Cees~H. Taal, Richard~C. Hendriks, Richard Heusdens, and Jesper Jensen.
\newblock A short-time objective intelligibility measure for time-frequency
  weighted noisy speech.
\newblock In \emph{2010 IEEE International Conference on Acoustics, Speech and
  Signal Processing}, pages 4214--4217, 2010.
\newblock \doi{10.1109/ICASSP.2010.5495701}.

\bibitem[Taal et~al.(2011)Taal, Hendriks, Heusdens, and Jensen]{Taal:11}
Cees~H. Taal, Richard~C. Hendriks, Richard Heusdens, and Jesper Jensen.
\newblock An algorithm for intelligibility prediction of time–frequency
  weighted noisy speech.
\newblock \emph{IEEE Transactions on Audio, Speech, and Language Processing},
  19\penalty0 (7):\penalty0 2125--2136, 2011.
\newblock \doi{10.1109/TASL.2011.2114881}.

\bibitem[Tan and Wang(2019)]{Wang:19}
Ke~Tan and DeLiang Wang.
\newblock Complex spectral mapping with a convolutional recurrent network for
  monaural speech enhancement.
\newblock In \emph{ICASSP 2019 - 2019 IEEE International Conference on
  Acoustics, Speech and Signal Processing (ICASSP)}, pages 6865--6869, 2019.
\newblock \doi{10.1109/ICASSP.2019.8682834}.

\bibitem[Tomar(2006)]{FFmpeg:06}
Suramya Tomar.
\newblock Converting video formats with ffmpeg.
\newblock \emph{Linux Journal}, 2006, 06 2006.

\bibitem[Tzanetakis and Cook(2002)]{GTZAN:02}
G.~Tzanetakis and P.~Cook.
\newblock Musical genre classification of audio signals.
\newblock \emph{IEEE Transactions on Speech and Audio Processing}, 10\penalty0
  (5):\penalty0 293--302, 2002.
\newblock \doi{10.1109/TSA.2002.800560}.

\bibitem[Wang(2021)]{pesq-github:21}
Wang.
\newblock {GitHub} - ludlows/python-pesq: {PESQ} ({Perceptual} {Evaluation} of
  {Speech} {Quality}) {Wrapper} for {Python} {Users} (narrow band and wide
  band), 2021.
\newblock URL \url{https://github.com/ludlows/python-pesq}.

\bibitem[Wang et~al.(2004)Wang, Bovik, Sheikh, and Simoncelli]{Wang:04}
Zhou Wang, Alan~C Bovik, Hamid~R Sheikh, and Eero~P Simoncelli.
\newblock Image quality assessment: from error visibility to structural
  similarity.
\newblock \emph{IEEE transactions on image processing}, 13\penalty0
  (4):\penalty0 600--612, 2004.

\end{thebibliography}

\end{document}